# Good Solid-State Electrolytes Have Low, Glass-like Thermal Conductivity

*Zhe Cheng, Beniamin Zahiri, Xiaoyang Ji, Chen Chen, Darshan Chalise, Paul V. Braun\*, David G. Cahill\**

Dr. Z. Cheng, Dr. B. Zahiri, X. Ji, C. Chen, Prof. P. V. Braun, Prof. D. G. Cahill

Department of Materials Science and Engineering, University of Illinois at Urbana-Champaign, Urbana, IL 61801, USA

Emails: pbraun@illinois.edu, d-cahill@illinois.edu

Dr. Z. Cheng, Dr. B. Zahiri, X. Ji, C. Chen, D. Chalise, Prof. P. V. Braun, Prof. D. G. Cahill

Materials Research Laboratory, University of Illinois at Urbana-Champaign, Urbana, IL 61801, USA

Dr. B. Zahiri, C. Chen, Prof. P. V. Braun

Beckman Institute for Advanced Science and Technology, University of Illinois at Urbana-Champaign, Urbana, IL 61801, USA

D. Chalise, Prof. D. G. Cahill

Department of Physics, University of Illinois at Urbana-Champaign, Urbana, IL 61801, USA

Prof. P. V. Braun

Department of Chemistry, University of Illinois at Urbana-Champaign, Urbana, IL 61801, USA

## Abstract

Management of heat during charging and discharging of Li-ion batteries is critical for their safety, reliability, and performance. Understanding the thermal conductivity of the materials comprising batteries is crucial for controlling the temperature and temperature distribution in batteries. This work provides systemic quantitative measurements of the thermal conductivity of three important classes of solid electrolytes (oxides, sulfides, and halides) over the temperature range $150 < T < 350$ K. Studies include the oxides $Li_{1.5}Al_{0.5}Ge_{1.5}(PO_4)_3$ and $Li_{6.4}La_3Zr_{1.4}Ta_{0.6}O_{12}$, sulfides $Li_2S$-$P_2S_5$, $Li_6PS_5Cl$, and $Na_3PS_4$, and halides $Li_3InCl_6$ and $Li_3YCl_6$. Thermal conductivities of sulfide and halide solid electrolytes are in the range 0.45-0.70 W m$^{-1}$ K$^{-1}$; thermal conductivities of $Li_{6.4}La_3Zr_{1.4}Ta_{0.6}O_{12}$ and $Li_{1.5}Al_{0.5}Ge_{1.5}(PO_4)_3$ are 1.4 W m$^{-1}$ K$^{-1}$ and 2.2 W m$^{-1}$ K$^{-1}$, respectively. For most of the solid electrolytes studied in this work, the thermal conductivity increases with increasing temperature; i.e., the thermal conductivity has a glass-like temperature dependence. The measured room-temperature thermal conductivities agree well with the calculated minimum thermal conductivities indicating the phonon mean-free-paths in these solid electrolytes are close to an atomic spacing. We attribute the low, glass-like thermal conductivity of the solid electrolytes investigated to the combination of their complex crystal structures and the atomic-scale disorder induced by the materials processing methods that are typically needed to produce high ionic conductivities.

# Introduction

Lithium-ion batteries (LIBs) are widely used for electrical energy storage. The kinetics of charge transport, intercalation, and chemical reactions[1, 2] in LIBs are strongly dependent on temperature. These processes, in turn, control cell round-trip efficiency and cycle life. For example, side reactions that irreversibly consume Li ions, e.g., electrolyte decomposition, are accelerated at high temperatures. Furthermore, exposure of a battery to elevated temperatures, while beneficial for fast charging, leads to significant safety risks due to the potential for electrolyte decomposition, oxygen release by cathode materials, and if heat cannot be removed at an appropriate rate, thermal runaway. Greater understanding of the thermal properties of battery materials will facilitate advances in the engineering design and thermal management of LIBs needed to improve safety and performance.

The thermal properties of materials used in liquid-electrolyte-based batteries were summarized by Kantharaj *et al*. at both the device and material levels.[2] An additional level of complexity is created by the fact that the thermal conductivity of cathodes and anodes are known to depend on the state of charge; e.g., the thermal conductivities of $LiCoO_2$ and graphite depend on Li composition.[3] From the cell assembly perspective, interfaces play a significant role in heat transfer.[4] For example, Lubner *et al.* characterized thermal resistances in a pouch-cell and showed that the thermal resistance of the separator-electrode interfaces accounts for up to 65% of the total thermal resistance within the cell.[4]

Solid-state batteries (SSBs) are under intense investigation as safer, more robust, and higher energy-density alternatives to liquid-electrolyte-based LIBs. Similar to their liquid-electrolyte-based counterparts, SSBs also require thermal management strategies for dissipation of heat, especially in large, high-energy cell stacks.[2, 5, 6] If lithium metal is used as the SSB anode, which is highly attractive due to lithium's low reduction potential (-3.04 V vs. standard-hydrogen-electrode) and high specific capacity (3860 mAh g$^{-1}$),[7] temperature control becomes particularly important, as control of internal temperature distributions enhances the homogeneity of Li deposition and suppresses the formation of dendrites.[6] In SSBs, the thermal properties of solid electrolytes (SEs) may play a key role in thermal management, in particular if SE thermal conductivities are low, in which case they will contribute disproportionately to the overall thermal resistance of the cell. However, to date, the thermal properties of only a few oxide and polymer solid electrolytes have been investigated.[8] Two emerging classes of SEs based on sulfide and halide chemistry possess high ionic conductivity—on the order of few mS cm$^{-1}$, comparable to liquid electrolytes—and stability at elevated temperature.[9] The thermal properties of halide and sulfide SEs are, however, essentially unknown, in part due to the experimental challenges created by their extreme sensitivity to air and moisture.

Here we report the temperature dependent thermal conductivity of some of the most important and emerging SEs as measured by time-domain thermoreflectance (TDTR). Oxides include $Li_{1.5}Al_{0.5}Ge_{1.5}(PO_4)_3$ and garnet $Li_{6.4}La_3Zr_{1.4}Ta_{0.6}O_{12}$; sulfides include Li conducting glassy $Li_2S$-$P_2S_5$, its halogenated crystalline derivative $Li_6PS_5Cl$, and Na-conducting cubic $Na_3PS_4$; the emerging family of halides include Li conductors $Li_3InCl_6$ and $Li_3YCl_6$. Our measurements of the thermal conductivity of air-sensitive samples (sulfides and halides) are enabled by deposition of the Al thin film transducer layer on the sample using a sputter deposition system that is housed within an Ar glove box. Thus, the samples are prepared, coated with an Al film, and mounted for TDTR measurements without exposure to air. X-ray diffraction (XRD) is used to characterize the crystal structure; differential scanning calorimetry (DSC) is used to measure the heat capacities. Finally, we compare the measured thermal conductivity at room temperature to the calculated minimum thermal conductivity and find good agreement.

**Methods and Samples**

We measured seven SE samples and two validation samples of Li salts, see **Table 1**. A single-crystal of LiF (purchased from MTI Corporation) is used as one of the validation sample because the thermal conductivity of LiF is well-known from prior experiments.[10, 11] Additionally, we studied a second validation sample of a pressed LiCl pellet formed from LiCl powders following the same procedure as the other SE samples. The thermal conductivity of LiCl has not, to the best of our knowledge, been reported previously; a first-principles calculation of the thermal conductivity of LiCl was recently published and provides a point of comparison.[12]

The Young's modulus and Poisson's ratio listed in Table 1 are used in the calculation of the minimum thermal conductivity. The relative density is calculated from the measured mass and volume of the pellet and dividing by the theoretical density based on lattice constants. The relative density of the SEs are greater than 96%. The thickness of the pellets, in the range of 300 to 500 µm was measured using a micrometer with accuracy of 1 µm. At least three thickness measurements were performed on each pellet and a range of 1-3% error was found for density values. The number of atoms ($n$) in the primitive cells of each SE material are calculated by converting from the conventional unit cells[13] with VESTA. The number of acoustic and optical phonon branches in the phonon dispersion is 3 acoustic phonon branches and 3$n$-3 optical phonon branches.

The LLZTO and LAGP discs were purchased from MSE Supplies. MSE Supplies fabricates the LLZTO and LAGP pellets by high temperature sintering (1200-1300 °C). Crystalline $Li_6PS_5Cl$ (LPSC) powder was purchased from NEI Corporation. We pressed the LPSC powder into pellets for thermal measurements.

Lithium halides and amorphous sulfide SE powders were synthesized at U. of Illinois through ball milling. Sample preparation and mixing for all lithium and sodium SEs described below were done in an Ar filled glovebox (< 0.1 ppm $O_2$ and $H_2O$). For lithium halide SEs, dry lithium chloride (LiCl, Sigma Aldrich) was mixed with yttrium chloride ($YCl_3$, Alfa Aesar) or indium chloride ($InCl_3$, Alfa Aesar) in the appropriate molar ratios (3:1) to obtain $Li_3YCl_6$ (LYC) or $Li_3InCl_6$ (LIC) mixtures, respectively. The mixtures were then manually ground using a mortar and pestle. In the case of LYC, excess $YCl_3$ powder (10 wt%) was added during the hand grinding step to accommodate for the loss of yttrium chloride due to adhesion to the milling vessel as also reported in other studies.[14, 15] The amount of excess $YCl_3$ (10 wt%) was determined by recursive XRD analysis of the final product to achieve phase pure LYC. In the case of LIC, excess $InCl_3$ was not necessary since that powder does not adhere to grinding and milling containers. We used X-ray fluorescence (XRF) (experimental details included below) to provide an additional check on the composition of the LYC and LIC powders. The Y:Cl ratio of the LYC powder is 5.7; i.e., the composition of the LYC powder has approximately 5% excess Y. The In:Cl ratio of the LIC powder is 6.1, i.e., stoichiometric within the experimental uncertainties of 1%. Variations in the composition between different samples of LIC and LYC were also on the order of 1%.

High energy ball milling (SPEX 8000 Mixer/Mill) was then used for the synthesis of the final product. A stainless-steel vial with a range of steel balls sizes (4 balls with diameter of 6.5 mm and two balls of diameter 9.5 mm, ball:powder weight ratio of approximately 10) was used for all SEs. The milling process was performed in an intermittent manner. This is to prevent caking when the powder mixture forms agglomerates at the corners of the milling jar. After an hour of milling, the milling jar was opened in the glovebox and caked portions were removed off the walls and mixed back into the milled powder. The milling was then continued for another hour. In a typical 5 g batch, the LYC powder was achieved after 2 to 3 hours of intermittent milling while LIC powder was obtained after one hour. An annealing step of 2-3 hour at 523 K under Ar was used to improve crystallinity.[16, 17] Amorphous $Li_2S-P_2S_5$ was prepared using high energy ball milling following a procedure described previously.[18] No post-treatment was used for this sample.

To form the $Na_3PS_4$ (NPS) SE, a 3:1 molar ratio of sodium sulphide ($Na_2S$, Sigma Aldrich) and phosphorus pentasulfide ($P_2S_5$, Sigma Aldrich) was hand-ground in a mortar as the first step. Milling times of 1 to 4 hours were studied and a 2-hour milling period was selected. The final highly conductive phase (cubic-NPS) was achieved after 2 hours of annealing at 543 K under Ar. (The ionic conductivity of these materials are reported separately[19]). Further annealing at T>543 K resulted in the lower ion conductivity tetragonal phase of NPS.[20]

Pellets of halides and sulfides SEs were fabricated in our laboratory by pressing 100-150 mg of SE powder between two quartz glass discs (McMaster-Carr Company) of 12.7 mm diameter and 1.58 mm thickness at room temperature in an Ar filled glovebox. The glass discs and powder SE were pressed between two Ti rods housed by a poly(aryl-ether-ether-ketone) (PEEK) mold. The inside diameter of the PEEK mold is 13 mm. The compression pressure was 360 MPa (~ 4 metric ton), see **Figure 1**(a).

Figure 1(b) schematically depicts the magnetron sputter deposition of the 80 nm Al layer that serves as the transducer for the TDTR measurements. Al layers were deposited using a custom-built deposition system in an Ar glovebox which eliminates air and moisture contamination of SE pellets. After Al film deposition, the Al-coated pellets were sealed using Kapton tape and silver paste (Figure 1(c)) without exposure to air. The sealed samples were then transferred to a temperature-controlled, vacuum compatible microscope stage (INSTEC HCP621G) (Figure 1(d)).

X-ray diffraction (XRD) analysis of the SE pellets was performed using a D8 ADVANCE Plus diffractometer with a Cu K$\alpha$ ($\lambda$=1.5418 Å) source in powder mode. All air-sensitive samples were prepared inside the glovebox ($O_2$ and $H_2O$ < 0.1 ppm) and mounted in a custom-designed holder and protected by a 7-8 µm layer of polyimide (Kapton) (Chemplex Industries Inc.). We selected the Kapton film to maximize X-ray transmission and minimize background. Reference structural data were taken from the inorganic crystal structure database (ICSD).[21]

Energy dispersive X-ray fluorescence (EDXRF) measurements of LIC and LYC samples were carried out in Shimadzu EDX-7000 spectrometer using X-ray beam from a Rh anode operated at an acceleration voltage of 50 kV. A white x-ray beam (5 mm in diameter) from the Rh source was incident on the sample through a 4 micron thick polypropylene window; the window protects the samples against exposure to air and moisture. Fluorescence X-rays were collected by a Si drift detector (SDD). PCEDX Navi software was used for determining the percentage composition of Cl and In or Y in the sample. The quantification involved comparison of the Cl K-alpha intensities and the intensities of the L lines of In or Y using the fundamental parameters method which takes into account fluorescence cross-sections and the expected absorption of the fluorescence X-rays.[22] We tested our approach on powders of $InCl_3$ and $YCl_3$ and found compositions within 1% of the expected 1:3 stoichiometry.

**Table 1**. Basic information of the measured samples. "Abbr." is abbreviation and "N/A" is "not applicable". $E$ is Young's modulus; $\sigma$ is the Poisson's ratio; $n$ is the number of atoms in a primitive unit cell. The Poisson's ratios and Young's moduli of the listed materials except LPS are values for the polycrystalline average while those of LPS are for amorphous materials.

| Samples | Abbr. | Theo. density (kg m$^{-3}$) | Measured density (kg m$^{-3}$) | $E$ (GPa) | $\sigma$ | $n$ |
|---|---|---|---|---|---|---|
| LiF | LiF | 2600[13] | N/A | 126[23] | 0.20[23] | 2 |
| LiCl | LiCl | 2070[13] | N/A | 57[23] | 0.22[23] | 2 |
| $3(Li_2S)$-$P_2S_5$ | LPS | 1940[13] | 1940 | 19[24] | 0.32[25] | N/A |
| $Li_6PS_5Cl$ | LPSC | 1870[13] | 1880 | 22[26] | 0.37[26] | 13 |
| $3(Na_2S)$-$P_2S_5$ | NPS | 2000[27] | 1950 | 18[25] | 0.34[25] | 16 |
| $Li_3YCl_6$ | LYC | 2450[13] | 2380 | 38[28] | 0.27[28] | 30 |
| $Li_3InCl_6$ | LIC | 2707[13] | 2680 | 38(b) | 0.27(b) | 10 |
| $Li_{6.4}La_3Zr_{1.4}Ta_{0.6}O_{12}$ | LLZTO | 5470(a) | 5310(a) | 147[29] | 0.24[29] | 93.6 |
| $Li_{1.5}Al_{0.5}Ge_{1.5}(PO_4)_3$ | LAGP | 3410(a) | 3270(a) | 144(c)[26] | 0.25(c)[26] | 37 |

a) Product information from the vendor.

b) No available data in the literature. We use the values of LIC.
c) No available data in the literature. We use the values of LiTi$_2$(PO$_4$)$_3$.

TDTR is a pump and probe technique capable of measuring thermal properties of both bulk and nanostructured materials.[30] A modulated pump beam heats the sample surface periodically while a delayed probe beam detects the temperature variation of the sample surface via thermoreflectance.[31] The signal picked by a photodetector and a lock-in amplifier is fitted with an analytical heat transfer solution of the sample structure to infer unknown parameters. Details about TDTR can be found in the literature.[32] In this work, we used a 10× objective lens with pump/probe spot size of 5.5 μm 1/e$^2$ intensity radius and a 9.3 MHz modulation frequency. Regions on the samples with reflectivity > 98% were selected for TDTR measurements. The Al thermal conductivity used in the TDTR data fitting are obtained by measuring the electrical conductivity with the four-point probe method and applying the Wiedemann-Franz law. LAGP and LLZTO are air-stable while the other SEs are air-sensitive. The Al layers were deposited by two different sputter systems. One sputter system is in the glovebox and the other is in air. The Al thermal conductivity for samples LAGP and LLZTO is 170 W m$^{-1}$ K$^{-1}$ while that for the other SEs is 130 W m$^{-1}$ K$^{-1}$ because of the quality difference of the Al deposited in the glovebox. The typical uncertainties of the thermal conductivity measured by TDTR are ±10%.

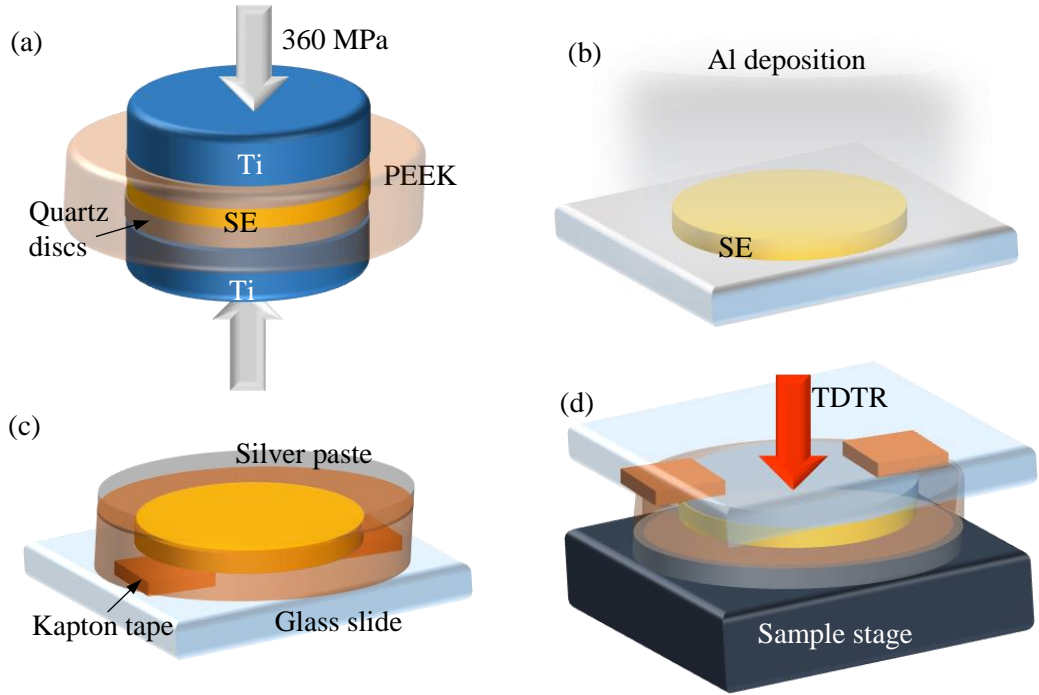

**Figure 1**. Schematic diagram of the sample fabrication process of the air-sensitive electrolytes. (a) Electrolyte powders (100-150 mg) are pressed into pellets with a pressure of 360 MPa. The outside diameter of the quartz discs (12.7 mm) is slightly smaller than the inside diameter of the PEEK holder (13 mm). (b) Al deposition in a custom-built sputtering system. (c) Sample sealing. The Al-coated side of the SEs face down. The small pieces of Kapton tape separate the SEs and the glass slide with a small gap. The edges of the SEs are sealed with Kapton tape and the top surface of the SEs are sealed with silver paste. (d) Samples are transferred to the TDTR sample stage in a vacuum chamber. Silver paste is used to attach the sample to the sample stage. The TDTR pump and probe laser beams are incident through the glass slide to reach the Al-coated SE samples.

To extract the thermal conductivity of SEs from the TDTR measurements, heat capacity data are required. Differential scanning calorimetry (DSC, Discovery 2500) was used to measure the heat capacity of the SEs. Since LAGP and LLZTO are air-stable, standard Al pans were used. The temperature was first increased to 423 K to remove any adsorbed water and increase sample-pan contact before running the DSC measurements. The other SEs were hermetically sealed in Al pans inside a glovebox. Prior to sealing, powders were pressed to the bottom of the pan using a 5-mm-diameter steel rod to improve contact. Before running the DSC measurements, the samples were heated to 398 K to increase sample-pan contact. The heating rate and the sample mass of all the DSC measurements are 10 K min$^{-1}$ and 10-20 mg, respectively.

**Results and Discussions**

**Figure 2** shows the XRD patterns of the six SEs that we studied that have crystalline structures: LLZTO, LAGP, LIC, LPSC, NPS, and LYC. Oxide SEs show the typical cubic structure of fully sintered oxide pellets as confirmed by cubic LLZTO and LAGP reference data. LPSC also resembles a typical cubic argyrodite structure in the $F\bar{4}3m$ space group in agreement with literature.[33] The structure of our NPS sample matches that of the cubic-NPS phase which is known to achieve the maximum sodium ion conductivity.[20] The XRD data for our LIC sample confirms that this sample possesses the distorted rock salt structure within the C2/m space group previously reported for similar ball-milled mixtures.[16] We also confirmed that the structure of our LYC sample matches the structure reported by Asano et al.[17] The XRD of amorphous LPS synthesized with the same method has been reported previously[18] and we did not repeat that measurement here.

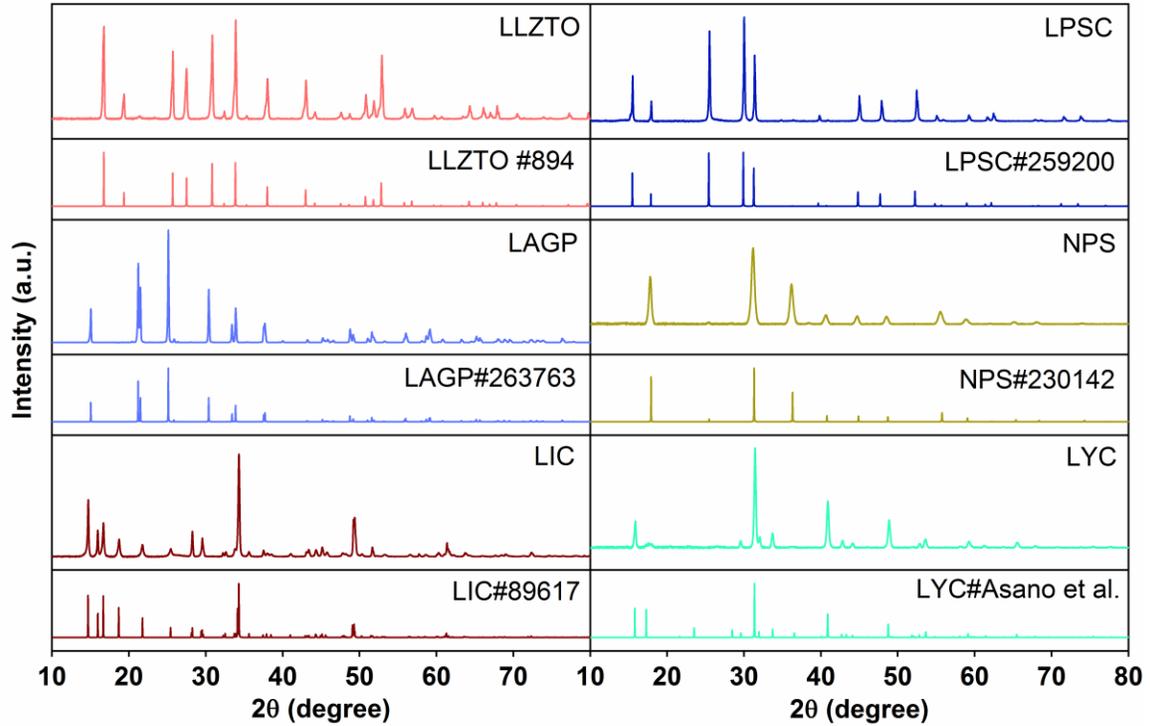

**Figure 2**. XRD patterns of crystalline SEs: LLZTO, LAGP, LIC, LPSC, NPS, and LYC. The XRD patterns of the same materials are in the same color. The upper pattern is the experimental data and the lower pattern is the reference structural data extracted from ICSD database.[21] The reference structural data of LYC is from the structure reported by Asano et al.[17]

**Figure 3**(a) shows the measured heat capacity of the air-stable oxide SEs: LLZTO and LAGP. The density used to calculate the volumetric heat capacity is the theoretical density listed in Table 1. As the temperature increases, the heat capacities of LLZTO and LAGP increase due to the increased number of thermally-excited phonons. The fact that the heat capacity is still increasing with increasing temperature at 400 K shows that the optical phonons of both LLZTO and LAGP are not fully thermally-excited at 400 K. For these oxide SEs, the optical phonon spectra extend to high frequencies because of the strong metal-oxygen bonds and the light mass of oxygen.

The heat capacities of sulfide SEs (LPS, NPS, and LPSC) and halide SEs (LIC and LYC) reach a nearly constant value at temperatures above room temperature (Figure 3(b)). Optical phonons in these materials are fully thermally-excited near room temperature because, in comparison to oxide SEs, the bonding is weaker and the mass of the anions is higher. Heat capacities of $Li_2S$ and $Na_2S$ calculated from first-principles are included for comparison.[34] At room temperature, the calculated heat capacities of $Li_2S$ and $Na_2S$ are 2.23 MJ m$^{-3}$ K$^{-1}$ and 1.66 MJ m$^{-3}$ K$^{-1}$, respectively, close to the classical limit based on the Dulong-Petit law (2.65 MJ m$^{-3}$ K$^{-1}$ and 1.76 MJ m$^{-3}$ K$^{-1}$). The experimental value for the heat capacity of NPS is close to the calculated value for $Na_2S$. Similarly, the experimental heat capacities of LPS and LPSC are close to the calculated values for $Li_2S$. A small peak appears in the heat capacity of LIC near 230 K. This peak is reproducible in multiple samples although the temperature of the peak varies between 220 K and 250 K. We do not yet understand the origin of this feature in the heat capacity of LIC.

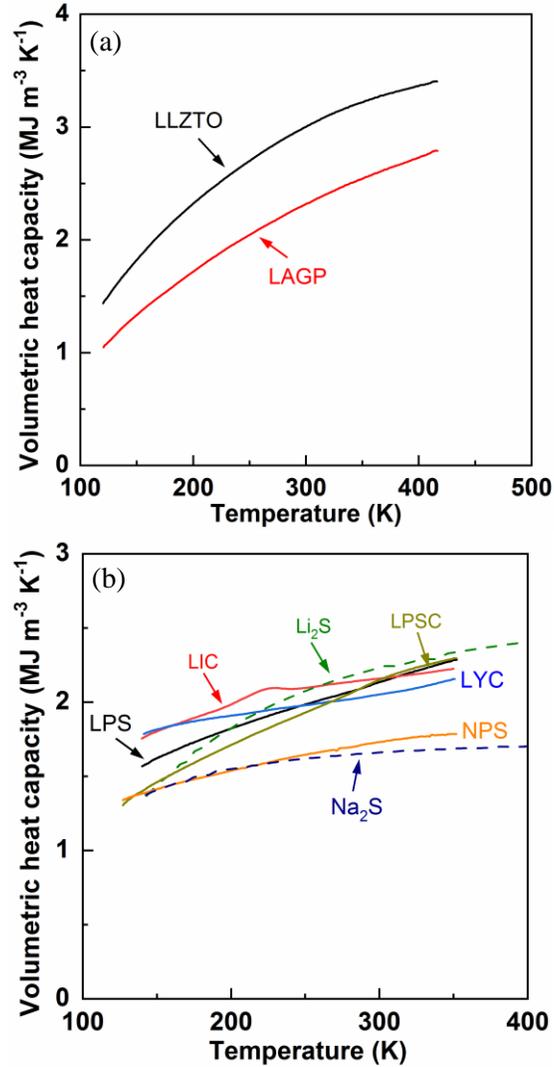

**Figure 3**. (a) DSC-measured heat capacity of air-stable oxide SEs: LAGP and LLZTO. (b) DSC-measured heat capacity of air-sensitive SEs: sulfides (LPS, LPSC, and NPS) and halides (LIC and LYC). Calculated values for $Li_2S$ and $Na_2S$ are shown as dashed lines and are included for comparison.[34]

The thermal conductivity of the two validation samples, a LiF single crystal and a LiCl pellet pressed from powders, are shown in **Figure 4**. Our measured thermal conductivity of LiF agrees well with experimental and calculated values in the literature.[10, 11] The thermal conductivity of LiF calculated from first-principles (black line) at high temperatures is proportional to the reciprocal of temperature due to the assumption that anharmonic interactions between three phonons are the dominant source of thermal resistance. The calculated thermal conductivity of LiF at room temperature by Xia *et al*. included both three-phonon and four-phonon scatterings.[10, 35]

The particle sizes of the LiCl powders are on the order of 100 microns, much larger than the TDTR laser spot size and the dominant phonon mean free path in LiCl. Details can be found in the Supporting Information. Spots on the LiCl samples with high reflectivity (≥98%) were selected to perform TDTR measurements. Figure 4(b) shows the measured temperature dependent thermal conductivity of LiCl pellets. The measured thermal conductivity matches the calculated thermal conductivity at room temperature.[12] The measured thermal conductivity is proportional to the reciprocal of temperature, as expected if the dominant source of thermal resistance is the intrinsic anharmonic interaction between three phonons. The agreement with the theoretically-predicted value and the temperature dependence support that the measured thermal conductivity is the intrinsic thermal conductivity of LiCl.

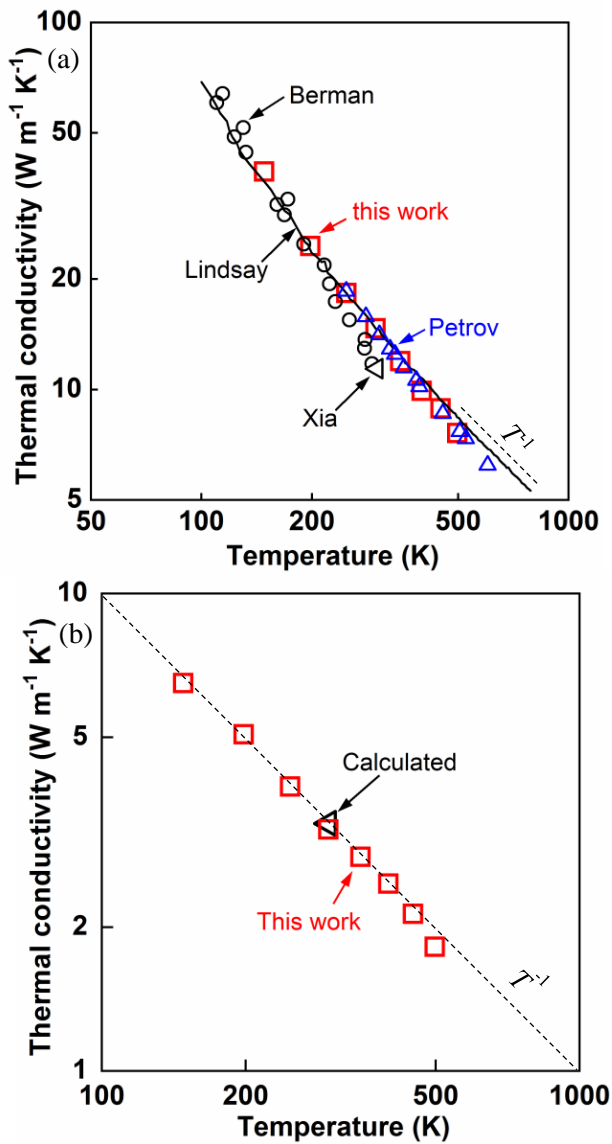

**Figure 4**. (a) Temperature dependent thermal conductivity of LiF. Open red squares are our measurements using TDTR. Prior measurements (Berman, open black circles; Petrov, open blue triangles) and calculated values (Lindsay, solid black line; Xia, open black triangle) are included for comparison.[10, 11, 35] (b) Temperature dependent thermal conductivity of LiCl. The calculated thermal conductivity of LiCl at room temperature is included for comparison.[12]

**Figure 5**(a) shows the temperature dependent thermal conductivity of the halide SEs. The thermal conductivity of LiCl is included for comparison. At room temperature, the thermal conductivity of LiCl is approximately 5 times larger than the thermal conductivity of LIC and LYC and has a fundamentally different temperature dependence. Across the temperature range of our measurements, 150 <T< 350 K, the thermal conductivities of LIC and LYC increase with increasing temperature. Similar temperature dependence of thermal conductivity has been previously observed in β-alumina SE and stabilized zirconia.[36, 37]

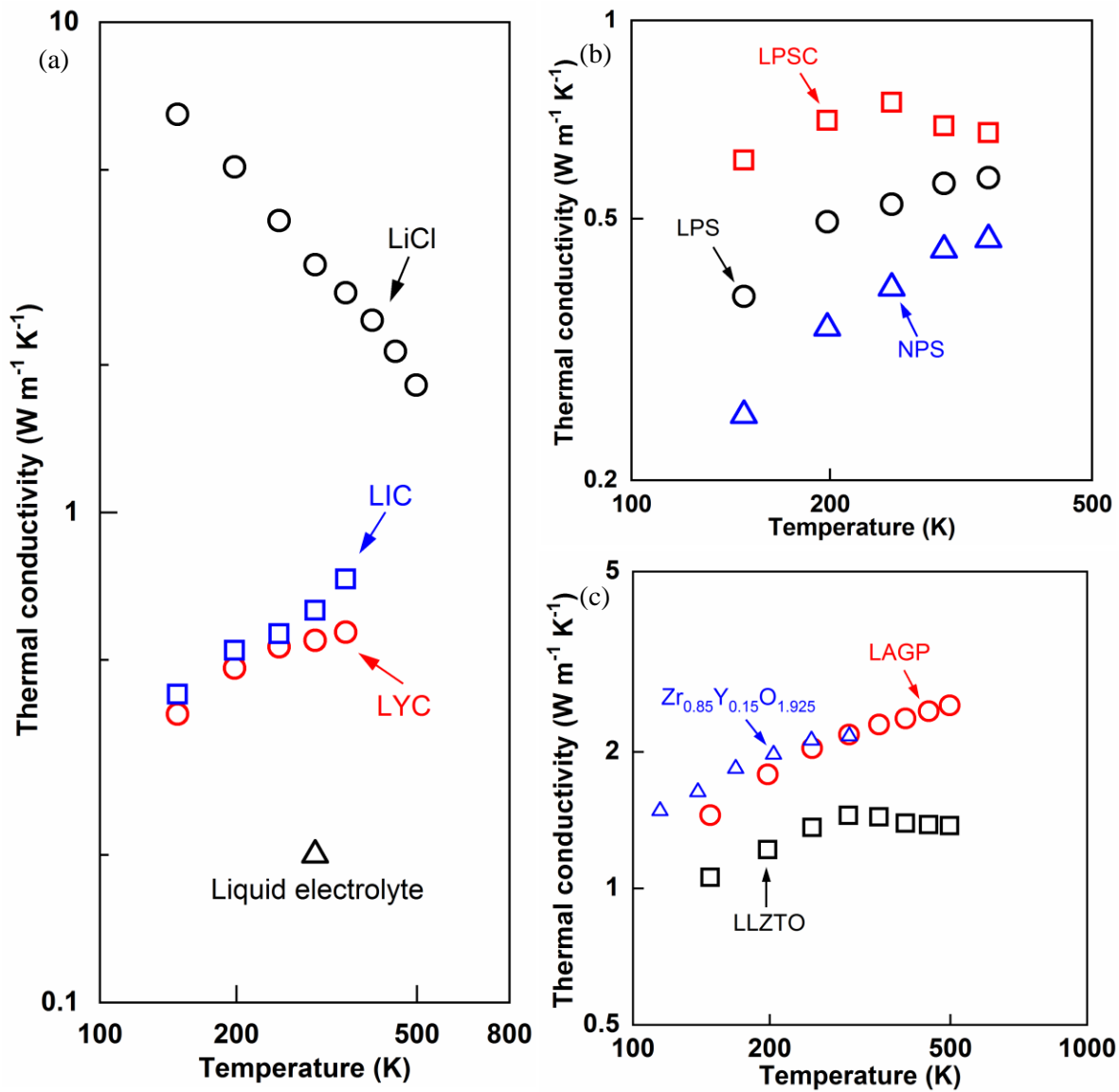

**Figure 5**. (a) Temperature dependent thermal conductivity of the halide SEs (LIC and LYC). The thermal conductivities of LiCl and liquid electrolyte are included for comparison.[4] The composition of the liquid electrolyte is 1M LiPF$_6$ salt concentration dissolved in 50 wt. % ethylene carbonate and 50 wt. % diethyl carbonate. (b) Temperature dependent thermal conductivity of the sulfide SEs (LPSC, LPS, and NPS). (c) Temperature dependent thermal conductivity of the oxide SEs (LLZTO and LAGP). The thermal conductivity of single crystal $Zr_{0.85}Y_{0.15}O_{1.925}$ is included for comparison.[37]

Figure 5(b) shows the temperature dependent thermal conductivity of sulfide SEs (LPSC, LPS, and NPS). The LPS that we studied is amorphous while the LPSC and NPS samples are crystalline. All the measured thermal conductivities are below 0.7 W m$^{-1}$ K$^{-1}$ from 150 K to 350 K. The thermal conductivity of LPS increases with increasing temperature, showing a typical temperature dependence for amorphous materials.[37, 38] A broad maximum appears in the thermal conductivity of LPSC near 250 K. NPS has the lowest thermal conductivity among all the SEs studied in this work and its thermal conductivity also shows a temperature dependence characteristic of amorphous materials.

Figure 5(c) shows the temperature dependent thermal conductivity of the oxide SEs (LLZTO and LAGP). The measured thermal conductivity is 1.4 W m$^{-1}$ K$^{-1}$ for LLZTO and 2.2 W m$^{-1}$ K$^{-1}$ for LAGP at room temperature, similar to amorphous oxides such as SiO$_2$ and Nb$_2$O$_5$.[37-39] The thermal conductivity of single crystal Zr$_{0.85}$Y$_{0.15}$O$_{1.925}$ is included for comparison.[37] We observed similar thermal conductivity for single crystal Zr$_{0.85}$Y$_{0.15}$O$_{1.925}$ and the LAGP SE in this work. The thermal conductivity of LAGP has an amorphous-like temperature trend. For the thermal conductivity of LLZTO, we observe a modest decrease at high temperatures.

**Figure 6** shows a comparison of the measured thermal conductivity of the SEs at room temperature with the corresponding calculated value of the minimum thermal conductivity for each material. The minimum thermal conductivity model in the high temperature limit is:[37]

$$\kappa_{\min} = \left(\frac{\pi}{48}\right)^{1/3} k_B n_a^{2/3} (v_l + 2v_t), \qquad (1)$$

$$v_l = \left[\frac{E(1-\sigma)}{(1+\sigma)(1-2\sigma)\rho}\right]^{1/2}, \qquad (2)$$

$$v_t = \left[\frac{E}{2(1+\sigma)\rho}\right]^{1/2}, \tag{3}$$

where $n_a$ is the number of atoms per unit volume; $k_B$ is the Boltzmann constant; $E$ is the Young's modulus; $\rho$ is the density; $\sigma$ is the Poisson's ratio; $v_l$ and $v_t$ are the longitudinal and transverse speed of sound, respectively. In the minimum thermal conductivity model, the lifetime of each Debye-like vibrational mode is assumed to be one half the period of vibration. As shown in Figure 6, the measured room-temperature thermal conductivity agrees well with the calculated minimum thermal conductivity, showing that the phonon mean free paths are comparable to the atomic spacing.

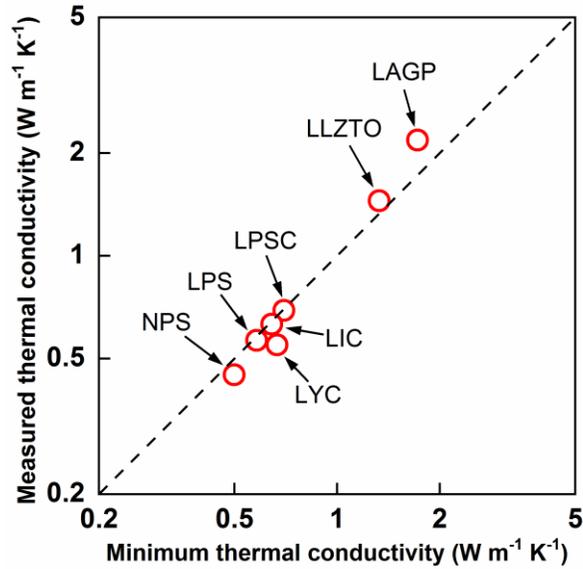

**Figure 6**. Comparison of the measured thermal conductivity of the SEs at room temperature with the calculated minimum thermal conductivity.

There are several mechanisms which alone or jointly can lead to glass-like thermal conductivities in crystalline materials: disorder, anharmonicity, and complex crystal structure.[37, 40, 41, 42] Disordered crystals such as single crystal $Zr_{0.85}Y_{0.15}O_{1.925}$ have glass-like thermal conductivity even though its fluorite crystal structure is not particularly complex.[37] Recent theoretical studies have revealed that the combination of strong anharmonicity and unit cell complexity can lead to glass-like thermal conductivity without disorder.[40, 41]

For the SE materials studied in this work, the large numbers of atoms in the primitive unit cells, as shown in Table 1, result in a large number of optical phonon branches ($3n-3$). Optical phonon branches typically have low phonon group velocities and contribute less to thermal conductivity than acoustic phonons. The relative population of acoustic phonons becomes relatively small if there are large numbers of optical phonon branches. As the increase in $n$, the ratio of acoustic branches to optical branches is $1/(n-1)$.

Additionally, the ball milling process generates defects such as cation site disorder.[15] The atomic-scale disorder induced by processing is typically necessary to produce high ionic conductivity. Annealing of ball-milled materials reduces the disorder and relaxes the structure into a more, but not fully, ordered structure.[15] Off-stoichiometric compositions also create atomic-scale disorder; for example, the sample of LLZTO we studied contains a random arrangement of Zr and Ta atoms.[29, 43] We tentatively conclude that disorder plays a primary role in producing the low, glass-like thermal conductivities because relatively modest amounts of disorder are known to produce glass-like thermal conductivity in crystals.[37] We cannot however rule out that a combination of strong anharmonicity and the complexity of the unit cell is also playing an important role in this behavior.[40] Further experimental tests of the physical mechanisms will require the synthesis SE crystalline materials with levels of atomic-scale disorder that can be varied over a wide range.

The development of SE with high thermal conductivity would potentially facilitate heat dissipation and improve thermal management of battery cells. A main conclusion of our work is that materials with both high thermal conductivity and high ionic conductivity are unlikely. A composite made of a good SE and a high thermal conductivity materials is a potential strategy to obtain enhanced thermal conductivity. The high thermal conductivity component would need to be chemically stable during the battery operation and electrically insulating, and not reduce the ionic conductivity of the composite electrolyte significantly. In addition to the intrinsic thermal resistance of the SE, the thermal contact resistance of the SE-electrode interfaces can also contribute significantly to the total thermal resistance of a battery cell.[4] The thermal contact resistance is not an intrinsic property of intimately bonded interfaces—for intimately bonded interfaces, this contact resistance would be negligible[44]—but instead arises from regions of physical separation between the SE and the electrode.

**Conclusions**

At room temperature, the measured thermal conductivity of the sulfide and halide SEs are below 0.7 W m$^{-1}$ K$^{-1}$ while the oxide SEs (LLZTO and LAGP) have thermal conductivity of 1.4 W m$^{-1}$ K$^{-1}$ and 2.2 W m$^{-1}$ K$^{-1}$, respectively. For most of the SEs studied in this work, the measured thermal conductivity increases with increasing temperature from 150 K to 350 K, showing a glass-like thermal conductivity trend with temperature. By calculating the minimum thermal conductivity, we observe excellent agreement with the measured room-temperature thermal conductivity. The measured thermal conductivity of SEs are several times higher than those of liquid electrolytes (0.2 W m$^{-1}$ K$^{-1}$), especially for the thermal conductivity of oxide SEs which are approximately one order of magnitude higher. Compared to liquid-electrolyte batteries, the higher thermal conductivity of SEs will facilitate heat dissipation in solid-state batteries. Additionally, the measured thermal conductivity of SEs will provide guidance for the selection of SEs for battery thermal management.

**Conflicts of Interests**: The authors claim no conflicts of interests.

**Supporting Information**
Supporting Information is available from the Wiley Online Library or from the author.

**Acknowledgements**
Z.C. and B. Z. contributed equally to this work. The authors acknowledge financial support from US Army CERL W9132T-19-2-0008. This research was carried out in part in the Materials Research Laboratory Central Research Facilities, University of Illinois.

Received: ((will be filled in by the editorial staff))
Revised: ((will be filled in by the editorial staff))
Published online: ((will be filled in by the editorial staff))